\documentstyle[aps]{revtex} 
\begin{document}
\twocolumn[
\title{Schwinger's Dynamical Casimir Effect: Bulk Energy Contribution}
\author{
C. E. Carlson$^{*}$, C. Molina--Par\'{\i}s$^{+}$,
J. P\'erez--Mercader$^{++}$, and Matt Visser$^{+++}$
}
\address{
$^{*}$Physics Department, College of William and Mary,
Williamsburg, Virginia 23187\\
$^{+}$Theoretical Division, Los Alamos National Laboratory,
Los Alamos, New Mexico 87545\\
$^{++}$Laboratorio de Astrof\'{\i}sica Espacial y F\'{\i}sica
Fundamental, Apartado 50727, 28080 Madrid\\
$^{+++}$Physics Department, Washington University,
St. Louis, Missouri 63130-4899\\
}
\date{23 September 1996}
\maketitle
\parshape=1 0.75in 5.5in \indent
{\small Schwinger's Dynamical Casimir Effect is one of several
candidate explanations for sonoluminescence.  Recently,  several
papers have claimed that Schwinger's estimate of the Casimir energy
involved is grossly inaccurate. In this letter, we show that  these
calculations omit the crucial volume term.  When the missing term
is correctly included one finds full agreement with Schwinger's
result for the Dynamical Casimir Effect. We have nothing new to
say about sonoluminescence itself except to affirm that the Casimir
effect is energetically adequate as a candidate explanation.}
]
\def\Re{{\rm Re}}
\def\Im{{\rm Im}}
\section{Introduction}

Several years ago, Schwinger wrote a series of
papers~\cite{Schwinger0,Schwinger1,Schwinger2} wherein he calculated
the Casimir energy released in the collapse of a spherically
symmetric bubble or cavity. Using general arguments, he showed that
the effect was mostly a volume effect, and derived a simple and
elegant formula for the energy release involved in the collapse.
He found that (for each polarization state) the {\em ``dielectric
energy, relative to the zero energy of the vacuum, [is given] by}

\begin{equation}
E = - V \int \frac{d^3\vec{k}}{(2 \pi)^3} \frac{1}{2} \,[\hbar c]\, k
\left( 1 -
\frac{1}{\sqrt{\epsilon}} \right).
\label{E-Schwinger-0}
\end{equation}

\noindent
{\em So the Casimir energy of a uniform dielectric is negative"}.
{}From the above one finds that a dielectric slab with a spherical
vacuum cavity of radius $R$ has a higher Casimir energy than the
same slab of material with the cavity re--filled with dielectric.
Introducing a wave-number cutoff $K$  into the previous
expression,~\footnote{For sonoluminescence, this wave-number cutoff
can be related to the wavelength of the electromagnetic radiation
emitted in the collapse of the bubble.  For generic dielectrics,
this wave-number cutoff is related to the high-wave-number asymptotic
behaviour of the dispersion relation.} and summing over polarization
states, shows that the Casimir energy of a cavity in a dielectric,
relative to pure dielectric, is

\begin{eqnarray}
E_{cavity}
&=&+2\frac{4\pi}{3}R^3 \; \int_0^K
{4 \pi k^2 dk\over(2\pi)^3}     \frac{1}{2}  \hbar c k
 \left(
 1 - \frac{1}{\sqrt{\epsilon}}
\right)
\nonumber\\
&=&+\frac{1}{6 \pi} \hbar c R^3 K^4
\left(1 - \frac{1}{\sqrt{\epsilon}} \right).
\end{eqnarray}

\noindent
In general, this volume term will be the dominant contribution.

In view of the elegance and simplicity of this result, it is natural
to ask whether it can also be derived by more traditional quantum
field theoretic means.  Indeed, the existence of such a volume
contribution is easy to verify on general physical grounds:

\noindent
(1) One can view Schwinger's result in elementary terms as simply
the difference in zero-point-energies, obtained by integrating the
difference in photon dispersion relations over the density of
states\footnote{The dots denote finite-volume corrections.
We shall develop this density-of-states point of view more fully
in a separate publication.}

\begin{equation}
E_{ cavity} = + 2 V \int \frac{d^3\vec{k}}{(2 \pi)^3} \frac{1}{2}
\hbar  \left[ c k - \omega(k)  \right] + \cdots
\end{equation}

\noindent
At low wave-numbers, we know that the dispersion relation for a
dielectric is simply summarized by the zero-frequency refractive
index $n$. That is
\begin{equation}
\omega(k) \to c k/n \qquad {\rm as} \qquad k \to 0.
\end{equation}
On the other hand, at high enough wave-numbers, the photons propagate
freely through the dielectric: They are then simply free photons
travelling through the empty vacuum between individual atoms. Thus

\begin{equation}
\omega(k) \to c k \qquad {\rm as} \qquad k \to \infty.
\end{equation}

{}From the above we know that the integrand must go to zero at
large wave-number.  In fact for any real dielectric the
integrand must go to zero sufficiently rapidly to make the integral
converge, since after all we are talking about a real physical
difference in energies. 

To actually calculate this energy difference one requires a suitable
physical model for $\omega(k)$.  Schwinger's calculation~\cite{Schwinger0},
is equivalent to picking the particularly simple model

\begin{equation}\label{omegamodel}
\omega(k) = {c k\over n} \; \Theta(K-k) + ck \; \Theta(k-K).
\label{eqn7}
\end{equation}

\noindent
Here $\Theta(x)$ is the Heaviside step function, and $K$ is a
wave-number which characterizes the transition from dielectric-like
behaviour to vacuum-like behaviour.  Note that the cutoff $K$
describes an actual physical situation: It is a surrogate for all
of the complicated physics that would be required to make a detailed
model for the dielectric to vacuum transition.

\noindent
(2) We also know that the quantum action in 3+1 dimensions generically
contains divergences which range from quartic to logarithmic, in
addition to finite contributions.  As is well known, this ``cosmological
constant" contribution (the quartic divergence) will not vanish
unless the theory has very special symmetries (for
example---super-symmetry). Thus energy densities that go as $({\rm
cutoff})^4$ are {\em generic} in (3+1) dimensions.

\noindent
(3) Alternatively, one could perform an explicit quantum field
theoretic calculation of the Casimir energy in some model problem
and thereby verify Schwinger's result.  A step in this direction
has been provided by Milton {\em et al.}~\cite{Milton80,Milton95,Milton96},
who attempted to compute the Casimir energy associated with a
spherical cavity of radius $R$, dielectric constant $\epsilon_1$,
and permeability $\mu_1$, embedded in an infinite medium with
dielectric constant $\epsilon_2$ and permeability $\mu_2$.   They
found that the dominant term is not the volume term but a surface
term which is proportional to $R^2 K^3 (\epsilon_1 -\epsilon_2)^2$.
We, however, have re-analysed these calculations and do find a
volume term which dominates except for very small bubbles.

We have performed the calculation of the Casimir energy in two
different  and complementary ways:

\noindent
(a) We have taken the formalism of Milton  {\em et
al.}~\cite{Milton80,Milton95,Milton96} and applied it directly to
an {\em ab initio} calculation of the Casimir energy. We compute
the energy difference between the following configurations:  (Case
I) an otherwise uniform medium with dielectric constant $\epsilon_2$
and permeability $\mu_2$, containing a spherical cavity of radius
$R$ with dielectric constant $\epsilon_1$ and permeability $\mu_1$,
and (Case II) a completely uniform medium with dielectric constant
$\epsilon_2$ and permeability $\mu_2$.  This energy difference is
given as a sum over a series of integrals involving Ricatti--Bessel
functions. Some of the sums can be evaluated explicitly while others
can only be evaluated by using an asymptotic analysis of the type
used by Milton {\em et al.} We verify the existence of both volume
and sub-dominant surface contributions.

\noindent
(b) We have analyzed the extant calculations to see exactly where
they differ from the present calculation. We find that the subtraction
scheme they use to calculate the Casimir energy does not correspond
to the physical situation in question. We isolate the difference
in energy between these calculations and the correct one. We will
explicitly show that this difference is proportional to volume.

\section{Physical description of the calculation}

Milton {\em et al.}~\cite{Milton80,Milton95,Milton96} explicitly
calculated the electromagnetic Green functions for a dielectric ball
embedded in an infinite space of (different) dielectric material, and
then attempted to calculate the Casimir energy by explicitly
integrating these Green functions over ``all space". Note that an
important limitation of any such calculation is that any attempt at
explicitly calculating Green functions must be restricted to systems
of extremely high symmetry---such as half-spaces, slabs, or balls.
The basic strategy is to take the classical expression for the energy

\begin{eqnarray}
E &=& {1\over2} \int_{Geometry}
    \left[ \epsilon \vec E^2 + {1\over\mu} \vec { B}^2 \right] d^3x,
\end{eqnarray}

\noindent
promote the electric and magnetic fields to be operator quantities,
and then calculate the vacuum expectation value

\begin{eqnarray}
E = {1\over2} \int_{Geometry}
    \big[ &\epsilon& \langle  \vec E(0,x) \cdot \vec E(0,x) \rangle
\nonumber\\
  +       &{1\over\mu}& \langle \vec { B}(0,x) \cdot \vec { B}(0,x)
                \rangle \big] d^3x.
\end{eqnarray}

\noindent
The geometry is incorporated in the calculation both via the limits
of integration and via the boundary conditions satisfied by the
fields.  Since these two-point functions are of course divergent,
they must be rendered finite by some regularization prescription.
Milton {\em et al.} use point-splitting in the time direction:

\begin{eqnarray}
E(\tau) = {1\over2} \int_{Geometry}
        \big[ &\epsilon& \langle \vec E(\tau,x) \cdot \vec E(0,x) \rangle
\nonumber\\
   +     &{1\over\mu}& \langle    \vec { B}(\tau,x) \cdot \vec { B}(0,x)
               \rangle ] d^3x.
\end{eqnarray}

All of the technical aspects of the analysis then focus on
the calculation of these two-point correlation functions (Green
functions) by explicitly solving for the TE and TM modes appropriate
for a spherical ball with dielectric boundary conditions; and then
explicitly writing down the Green functions as a sum over suitable
combinations of Ricatti--Bessel functions and vector spherical
harmonics. To avoid unnecessary notational complications, we
schematically rewrite the above as

\begin{equation}
E(\tau) = {1\over2} \int_{Geometry} G_{[\epsilon,\mu]}(\tau,x;0,x)
\; d^3x,
\end{equation}

\noindent
where $G_{[\epsilon,\mu]}(t,x;t',x')$ is simply shorthand for the
linear combination of Green functions appearing above.

We may calculate these Green functions for {\em three}
different geometries\footnote{Notice that in Milton {\it et
al.} the dielectric properties of these media are taken to be
frequency independent, the cutoff being put in ``by hand" via
time-splitting.}: \\
{\bf Case I:} A dielectric ball of dielectric constant $\epsilon_1$,
permeability $\mu_1$, and radius $R$ embedded in a infinite dielectric
of {\em different} dielectric constant $\epsilon_2$ and permeability
$\mu_2$. (In applications to sonoluminescence, think of this as an air
bubble of radius $R$ in water.)  \\
{\bf Case II:} A completely homogeneous space completely filled with
dielectric $(\epsilon_2,\mu_2)$. (In applications to sonoluminescence,
think of this as pure water.)  \\
{\bf Case III:} A completely homogeneous space completely filled with
dielectric $(\epsilon_1,\mu_1)$. (In applications to sonoluminescence,
think of this as pure air.)\\

We are in complete agreement with the extant calculations and
results for these three individual Green functions---where we
disagree, as will be shown, is {\em in the way that these three
Green functions are inserted into the computation for the Casimir
energy.}

Milton {\em et al.} calculate an ``energy difference", which we will
call $E_{surface}$, and which they define as
\begin{equation}
E_{surface} = {1\over2} \left\{ \int_{ \small all\ r} G_I
   - \int_{r>R} G_{II}  - \int_{r<R} G_{III} \right\}.
\end{equation}
The computation of this quantity in~\cite{Milton80,Milton95,Milton96}
is mathematically correct.  An asymptotic analysis shows that this
expression is indeed proportional to the surface area---plus even
higher-order terms. However, the energy calculated from the
above expression does not correspond to the energy of a physically
realizable situation.

The physically correct quantity to compute is~\cite{Schwinger0}
\begin{equation}
\label{E-casimir-define}
E_{Casimir} = {1\over2} \left\{ \int_{\small all\ r} G_I
                              - \int_{\small all\ r} G_{II} \right\}.
\end{equation}
Observe that this quantity is simply the difference in energy
between two real physical situations: (Case I) having the dielectric
ball present and (Case II) replacing the dielectric ball by the
surrounding medium. {\em This is exactly the quantity that Schwinger
calculates in reference~\cite{Schwinger0} to describe the Casimir
enegy released in the collapse of the bubble: it is the energy
released in evolving from bubble to no--bubble.} The difference
between the two calculations is

\begin{equation}
\Delta E
= E_{Casimir} - E_{surface}
= {1\over2} \int_{r<R} \left\{ G_{III} - G_{II} \right\}.
\end{equation}

\noindent
This difference is easily seen to be proportional to the volume:
remember that $G_{III}$ and $G_{II}$ are Green functions corresponding
to two spaces that are completely filled with homogeneous
dielectrics---therefore they are each individually translation
invariant. (When one expresses these Green functions in terms of
spherical polar coordinates this is not obvious.) This observation
permits one to pull the Green functions outside the integral, so
that
\begin{equation}
\Delta E = {1\over 2} V \left\{ G_{III}(\tau,0;0,0) -
G_{II}(\tau,0;0,0) \right\},
\end{equation}
\noindent
where $V$ is the volume of the ball of radius $R$.  We shall now
show that this term is in fact exactly in conformity with Schwinger's
result.

\section{ ``Ab initio" calculation}

\subsection{The energy density}

We now calculate the Casimir energy for the geometrical configuration
previously described. We use techniques developed by Milton {\em
et al.}, but use, as Schwinger did, a wave number cutoff and shall
present the calculation in as much detail as space permits.  We
defer many technical details to a forthcoming publication.

For each individual geometry, the energy density ${ T}^{ tt}$ can
be evaluated by using the dyadic Green function
formalism~\cite{Milton80,Milton95,Milton96}. (Henceforth we use
natural units.) For Case I one finds

\begin{equation}\label{E-energy-density}
T^{tt}_I(r) =
\Re \left[{-i\over8\pi} \int_{-\infty}^{+\infty} {d\omega\over 2\pi}
e^{-i\omega\tau} \; X_I(k,r)\right],
\end{equation}

\noindent
with identical expressions holding for the other geometries.  Here the
$\omega$--integral arises from the time-splitting regularization,
while we have used the notation $k = |\omega| n$ with $n$ the
appropriate position--dependent refractive index { ($n_1$ inside the
dielectric sphere, $n_2$ outside)}, and have defined the quantity
$X_I(k,r)$ by

\begin{eqnarray}
X_I(k,r) &\equiv&
\sum_{\ell=1}^\infty (2\ell+1)
\Bigg\{ \left[ k_{I}^2 + {\ell(\ell+1)\over r^2} \right]
F^{I}_\ell(k;r,r)
\nonumber\\
&&+ {1\over r^2} {\partial\over\partial r_1} r_1
{\partial\over\partial r_2} r_2
 \left[ F^{I}_\ell(k;r_1,r_2)  \right]\Big|_{r_1=r_2=r}
\Bigg\}
\nonumber\\
&&+  \left[ (F^I_\ell) \to (G^I_\ell) \right].
\end{eqnarray}

\noindent
The functions $F_\ell (r,r')$ and $G_\ell (r,r')$ are the Green
functions for the electrical and magnetic fields in the appropriate
geometry, and are given below.  Similar results, with
appropriate substitutions for the momenta, hold when one makes
reference to Cases II and III.  Note that when making the substitutions
$(I)\to(II)$ or $(I)\to(III)$, one should also change the refractive
index that implicitly appears in the factor $k$. In addition it
should be borne in mind that $k_I$ is a function of position: $k_I
= n_1 |\omega|= k_{III}$ inside the dielectric sphere, whereas $k_I
= n_2 |\omega| = k_{II}$ outside the dielectric sphere.

Because  $F_\ell$ and $G_\ell$ depend only on the {\em absolute}
value of $\omega$ we can write the energy density as

\begin{eqnarray}
T^{tt}_I(r)
&=&
\Re \left[ {-i\over8\pi}  \int_{0}^{\infty}
{d\omega\over 2\pi}
\left[ e^{-i\omega\tau} + e^{+i\omega\tau} \right]  X_I(k,r) \right]
\nonumber\\
&=&
\Re \left[{-i\over4\pi}   \int_{0}^{\infty}
{d\omega\over 2\pi}
\cos(\omega\tau) \;  X_I(k,r) \right]
\nonumber\\
&=&
{1\over4\pi} \int_{0}^{\infty}
{d\omega\over 2\pi}
\cos(\omega\tau) \; \Im\left[X_I(k,r)\right].
\label{E-density-formula}
\end{eqnarray}

\noindent
The Casimir energy, Eq. (\ref{E-casimir-define}) is obtained by
taking the {\em difference} in energy densities and integrating
over all space while paying attention to the appropriate index of
refraction for each region of space. Thus

\begin{eqnarray}
\label{E-casimir-energy-1}
E_{Casimir} &=& \int_0^\infty r^2 dr \;
\int_0^{\infty} {d\omega\over 2\pi}
\cos(\omega\tau)
\nonumber\\
&& \times \Im\left[X_I(k,r)-X_{II}(k,r)\right].
\end{eqnarray}

This expression for the Casimir energy is completely equivalent to
equation (41) of~\cite{Milton95}, and equation (4.2b) of~\cite{Milton96}
and is also closely related to equations (30a) and (30b)
of~\cite{Milton80}. Note that extant calculations use the same
time-splitting parameter for the two different media---the physics
behind this choice is far from clear, and we shall return to this
point in a future publication.

It is now clear how one should modify these expressions to replace
time-splitting regularization by a wave-number cutoff.  For generality
we can take an arbitrary wave-number cutoff described by some smooth
real function $f(k)$ which goes to zero as $k \rightarrow
\infty$ and simply write
\begin{eqnarray}
\label{E-casimir-energy-2}
E_{Casimir} &=& \int_0^\infty r^2 dr \;
\int_0^\infty {d\omega\over 2\pi}
\nonumber\\
&& \times
\Im[f(k_I) X_I(k,r)- f(k_{II}) X_{II}(k,r)].
\end{eqnarray}
With due caution, the relevant Green functions can be read
off from~\cite{Milton80,Milton95,Milton96}.

\subsection{The Green functions}

In evaluating the Green functions one must be careful to correctly
incorporate the boundary conditions appropriate to the geometry
and the physics. This means that they must satisfy appropriate
continuity conditions derived from Maxwell's equations. That is

\begin{equation}
\vec E_\perp, \quad \epsilon \vec E_r, \quad
{1\over\mu}\vec B_\perp, \quad {\rm and} \quad \vec B_r,
\end{equation}

\noindent
must be continuous. In terms of $F_{\ell}$ and $G_{\ell}$,
one sees that

\begin{equation}
\mu F_{\ell}, \quad G_{\ell}, \quad
{\partial \over \partial r}{r F_{\ell}}, \quad {\rm and} \quad
\frac{1}{\epsilon}{\partial \over \partial r}{r G_{\ell}},
\end{equation}

\noindent
must be continuous. The Green functions are

\noindent
{\bf Case I:} \\
For $r_1, r_2 < R$:
\begin{eqnarray}
F^I_\ell, G^I_\ell(r_1,r_2) &=&
i k_{III} \; j_\ell(k_{III} r_<)
\nonumber\\
&&\times
\left[h_\ell(k_{III}r_>) - A^\ell_{F,G} \; j_\ell(k_{III} r_>) \right].
\label{E-i-a}
\end{eqnarray}
For $r_1, r_2 >R$:
\begin{eqnarray}
F^I_\ell, G^I_\ell(r_1,r_2) &=&
i k_{II} \; h_\ell(k_{II} r_>)
\nonumber\\
&&\times
\left[j_\ell(k_{II}r_<) - B^\ell_{F,G} \; h_\ell(k_{II} r_<) \right].
\label{E-i-b}
\end{eqnarray}

\noindent
The function $j_\ell(x)$ is the spherical Bessel function of order
$\ell$ and $h_\ell(x) \equiv h_\ell^{(1)} (x)$ is the spherical
Hankel function of the first kind. See equations (12a) and (12b)
of~\cite{Milton80},  equation (16) of~\cite{Milton95}, or equation
(2.13) of~\cite{Milton96}.  The quantities $A^\ell_{F,G}$ and
$B^\ell_{F,G}$ are those given in~\cite{Milton80,Milton95,Milton96}.

\noindent
{\bf Case II:}\\
 For all $r_1,r_2$:
\begin{equation}
F^{II}_\ell, G^{II}_\ell(r_1,r_2)
= i k_{II} \; j_\ell(k_{II} r_<)   \; h_\ell(k_{II}r_>).
\label{E-ii}
\end{equation}

\noindent
{\bf Case III:}\\
For all $r_1, r_2$:
\begin{equation}
F^{III}_\ell, G^{III}_\ell(r_1,r_2)
= i k_{III} \; j_\ell(k_{III} r_<) \; h_\ell(k_{III}r_>).
\label{E-iii}
\end{equation}

\noindent
We are now ready to explicitly compute the Casimir energy. In
passing we remark that the object $F^{(0)}_\ell$ defined in equation
(29) of~\cite{Milton80}, equation (35) of~\cite{Milton95}, and
equation (3.7) of~\cite{Milton96}, which is essential for those
calculations, is {\em not} a Green function of {\em any} differential
operator. Specifically, $F^{(0)}_\ell$ does not satisfy the dielectric
boundary conditions.  It is not even continuous, and is merely a
potpourri of two different Green functions which does not have any
particular physical relevance.

\subsection{The Casimir energy}

We calculate $E_{Casimir}$ using equation (\ref{E-casimir-energy-2}),
[equivalently (\ref{E-casimir-energy-1})] together with equations
(\ref{E-i-a}--\ref{E-ii}).  When evaluating the imaginary parts of
$X$ it is convenient to introduce the Ricatti--Bessel functions
$s_\ell(x) = { x j}_\ell(x)$ and $e_\ell(x) = { x h}_\ell(x)$.
One also needs the identity

\begin{equation}
{\ell(\ell+1)\over x^2} s_\ell(x) = s_\ell''(x) + s_\ell(x),
\end{equation}

\noindent
together with an identical equation which holds for
$e_\ell(x)$. After some rearrangement (technical details are
suppressed and will be relegated to a more detailed forthcoming
publication) we find that inside the dielectric sphere

\begin{eqnarray}
\Im\{&X_I&(k,r)\}_{in}
\nonumber\\
&=&
2\; {k_{III}\over r^2} \; \sum_{\ell=1}^\infty (2\ell+1)
\nonumber\\
&&\times
\left\{ 2 [s_\ell(x)]^2 + [s_\ell'(x)]^2  + s_\ell(x) s_\ell''(x)
\right\}|_{k_{III}r}
\nonumber\\
&-&  {k_{III}\over r^2} \; \sum_{\ell=1}^\infty (2\ell+1)
\Re\{A^\ell_F+A^\ell_G
\}
\nonumber\\
&&\times
\left\{ 2 [s_\ell(x)]^2 + [s_\ell'(x)]^2  + s_\ell(x) s_\ell''(x)
\right\}|_{k_{III}r} .
\nonumber\\
\end{eqnarray}

A remarkable Ricatti--Bessel function identity permits us to perform
the first sum over $\ell$ {\em exactly}. Using
\begin{equation}
\sum_{\ell=1}^{\infty} (2\ell+1)
\left[ 2 s_\ell(x)^2 + s_\ell'(x)^2 + s_\ell(x)  s_\ell''(x) \right] =
2 x^2,
\end{equation}
one obtains that

\begin{eqnarray}
\Im\{ &X_I&(k,r)\}_{in} =4 k_{III}^3
\nonumber\\
&-&  {k_{III}\over r^2} \; \sum_{\ell=1}^\infty (2\ell+1)
\Re\{A^\ell_F+A^\ell_G
\}
\nonumber\\
&&\times
\left\{ 2 [s_\ell(x)]^2 + [s_\ell'(x)]^2  + s_\ell(x) s_\ell''(x)
\right\}\big|_{k_{III}r}.
\nonumber\\
\end{eqnarray}
Note that for the $A^\ell_{F,G}$--terms we {\em cannot} explicitly
perform the $\ell$ summation because of the complicated $\ell$
dependence of these coefficients~\cite{Milton95,Milton96}.

It is very important to notice at this point that the
$A^\ell_{F,G}$--terms in the above expressions are the {\em only}
pieces retained in the currently extant calculations, the
other terms unfortunately have been missed there due to the use of
the wrong ``Green functions".

Taking a cue from the above, the results for the region inside the
bubble can be written (using self explanatory notation) as

\begin{equation}
\Im\{ X_I(k,r) \}_{in}
= 4 n_1^3 |\omega|^3 + Q_{in}^{surface}(k_{III},r).
\end{equation}

Turning to the region outside the dielectric sphere, one gets that

\begin{eqnarray}
\Im\{ &X_I&(k,r)\}_{out}
\nonumber\\
&=& 2\; {k_{II}\over r^2} \; \sum_{\ell=1}^\infty (2\ell+1)
\nonumber\\
&&\times
\left\{ 2 [s_\ell(x)]^2 + [s_\ell'(x)]^2 + s_\ell(x)
s_\ell''(x)\right\}|_{k_{II}r}
\nonumber\\
&-& {k_{II}\over r^2} \; \sum_{\ell=1}^\infty (2\ell+1)
 \Re\big[ (B^\ell_F+B^\ell_G)
\nonumber\\
&&\times
\left\{ 2 [e_\ell(x)]^2 + [e_\ell'(x)]^2 + e_\ell(x) e_\ell''(x)
\right\}\big]\big|_{k_{III}r}
\nonumber\\
&=& 4 k_{II}^3 - {k_{II}\over r^2} \; \sum_{\ell=1}^\infty (2\ell+1)
 \Re\big[ (B^\ell_F+B^\ell_G)
\nonumber\\
&&\times
\left\{ 2 [e_\ell(x)]^2 + [e_\ell'(x)]^2 + e_\ell(x) e_\ell''(x)
\right\}\big]\big|_{k_{III}r} .
\end{eqnarray}

Again, in self-explanatory notation

\begin{equation}
\Im\{ X_I(k,r) \}_{out}
= 4 n_2^3 |\omega|^3 + Q_{out}^{surface}(k_{II},r).
\end{equation}

For Case II one simply has

\begin{eqnarray}
\Im\{ &X_{II}&(k,r) \}_{all~space}
=
2\; {k_{II}\over r^2} \; \sum_{\ell=1}^\infty (2\ell+1)
\nonumber\\
&&\times
\left\{ 2 [s_\ell(x)]^2 + [s_\ell'(x)]^2 + s_\ell(x)
s_\ell''(x)\right\}|_{k_{II}r}
\nonumber\\
&=& 4 k_{II}^3 = 4 n_2^3 |\omega|^3.
\end{eqnarray}

Going back to the momentum-space regulated Casimir energy, equation
(\ref{E-casimir-energy-2}), we obtain

\begin{eqnarray}
E_{Casimir}
&=& \int_0^R r^2 dr \int_{0}^{\infty} {d\omega \over2\pi}
4 |\omega|^3
\nonumber\\
&&\qquad \times\left[ n_1^3 f(n_1 |\omega|) - n_2^3 f(n_2
|\omega|)\right]
\nonumber\\
&+& \int_0^R r^2 dr \int_{0}^{\infty} {d\omega\over2\pi}
f(n_1 |\omega|) Q^{surface}_{in}(k,r)
\nonumber\\
&+& \int_R^\infty r^2 dr \int_{0}^{\infty} {d\omega\over2\pi}
f(n_2 |\omega|) Q^{surface}_{out}(k,r).
\nonumber\\
\end{eqnarray}

The remaining integrals for the $n^3 |\omega|^3$ term are trivial.
Changing the integration variable to $k = n |\omega|$, and
explicitly re-inserting the appropriate factors of $\hbar$ and $c$,
we get

\begin{eqnarray}
E_{Casimir}
&=&
+ 2 V \int {d^3\vec k \over (2\pi)^3} {1\over 2}
  \hbar [ \omega_1(k) - \omega_2(k) ] f(k)
\nonumber\\
&+& \int_0^R r^2 dr \int_{0}^{\infty} {d\omega\over2\pi}
\hbar f(n_1 |\omega|) Q^{surface}_{in}(k,r)
\nonumber\\
&+& \int_R^\infty r^2 dr \int_{0}^{\infty} {d\omega\over2\pi}
\hbar f(n_2 |\omega|) Q^{surface}_{out}(k,r).
\nonumber\\
\end{eqnarray}

\noindent
which is the central result of this paper.

This is explicitly of the form:

\begin{center}
(Schwinger's volume term) + (surface term).
\end{center}

\noindent
The ``surface term" corresponds to $E_{surface}$ and is given by the
two double integrals in the expression for $E_{Casimir}$ above. For
time-splitting regularization and dilute dielectric media, these
terms were explicitly shown by Milton {\em et al.}  to be
proportional to the surface area (plus even higher-order corrections).
The first term is the volume term 
not present in some of the existing calculations. In fact, after
approximating air by vacuum (setting
$n_1=1$) and using Schwinger's momentum space cutoff, this integral
is {\em exactly equal} to Schwinger's result~\cite{Schwinger0}.

\section{ Discussion}

The main result of this paper can be succinctly stated:  in a
dielectric medium of dielectric constant $n$ the Casimir energy
of a cavity---the difference in zero point energies of a
dielectric medium of refractive index $n$ with and without a vacuum
cavity of volume $V$---is:

\begin{equation}
E_{Casimir} =  {1\over8\pi^2} \; V \;  \hbar c  K^4 \;
\left[1-{1\over n}\right] + \cdots,
\end{equation}
with this volume term dominant if the scale of the bubble is {\em
larger} than the cutoff wavelength $2 \pi / K$.  This result is
completely in agreement with Schwinger's calculation in~\cite{Schwinger0},
and Schwinger's argument is now buttressed by our explicit
re-assessment of Milton {\em et al.}'s calculation for a spherical
dielectric ball.
 
We close with what is perhaps a minor point that we nevertheless
feel should be made explicit:  the volume contribution to the
Casimir energy is always there, and is always physical, but it is
{\em sometimes} safe to neglect it.

For example, a situation equally physical as the one we have
considered here is the following: suppose one is provided with a
fixed number of dielectric bodies of fixed shape (in particular,
of fixed volume), and suppose that one simply wishes to move the
bodies around in space with respect to each other. Then the bulk
volume contributions to the Casimir energy, while still present,
are constants independent of the relative physical location of the
dielectric bodies, and so merely provide a constant offset to the
total Casimir energy. {\em If all we are interested in is the energy
differences between different spatial configurations of the same
bodies then the various volume contributions can be quietly
neglected.}

On the other hand, the volume contribution is of critical importance
whenever one wants to calculate the energy difference between an
inhomogeneous dielectric and a homogeneous dielectric wherein the
irregularities have been filled in. {\em This is, precisely, the
physical situation in the case of bubble collapse in a dielectric
medium.}

\acknowledgements

This work was supported in part by the U.S. Department of Energy,
the U.S. National Science Foundation, by the Spanish Ministry of
Education and Science and the Spanish Ministry of Defense. Part of
this work was carried out at the Laboratory for Space Astrophysics
and Fundamental Physics (LAEFF, Madrid), and C.E.C., C.M-P., and
M.V. wish to gratefully acknowledge the hospitality shown. Part of
this work was carried out at Los Alamos National Laboratory, and
J.P-M. wishes to gratefully acknowledge the hospitality shown to
him there.


\end{document}